\documentstyle [12pt,a4p,epsfig,amsmath,multicol]{article}
\begin{document}
\vspace*{0.6cm}

\begin{center} 
{\normalsize\bf Absolute simultaneity and invariant lengths: Special Relativity 
 without light signals or synchronised clocks}
\end{center}
\vspace*{0.6cm}
\centerline{\footnotesize J.H.Field}
\baselineskip=13pt
\centerline{\footnotesize\it D\'{e}partement de Physique Nucl\'{e}aire et 
 Corpusculaire, Universit\'{e} de Gen\`{e}ve}
\baselineskip=12pt
\centerline{\footnotesize\it 24, quai Ernest-Ansermet CH-1211Gen\`{e}ve 4. }
\centerline{\footnotesize E-mail: john.field@cern.ch}
\baselineskip=13pt
\vspace*{0.9cm}
\abstract{It is demonstrated that
 the measured spatial separation of two objects, at rest in some inertial frame, is invariant under
 space-time transformations.
 This result holds in both Galilean and Special Relativity. A corollary is that there are no
 `length contraction' or associated 'relativity of simultaneity' effects in the latter theory.
  A thought experiment employing four unsynchronised clocks and a single measuring rod 
  reveals that the physical basis of the time dilatation effect is a relative velocity transformation
  law, not  `length contraction'. Time dilatation, which is universal and translation
   invariant for all synchronised clocks at rest in any inertial frame, is the
   unique space-time phenomenon discriminating Special from Galilean Relativity.}
 \par \underline{PACS 03.30.+p}
\vspace*{0.9cm}
\normalsize\baselineskip=15pt
\setcounter{footnote}{0}
\renewcommand{\thefootnote}{\alph{footnote}}

   \par In the conventional presentation of Special Relativity Theory (SRT), following 
   the approach of Einstein's seminal paper on the subject~\cite{Ein1}, light
   signals and clock synchronisation play a crucial role in the development of
   predictions of space-time geometrical effects on the basis of the Lorentz
   transformation (LT). However, as early as 1910~\cite{Ignatowsky},
 it was realised
   that Einstein's methodology is not the only, and not the most economical, way
   (in terms both of the number of postulates and their simplicity) to formulate the
    foundations of SRT. 
    \par In a previous paper by the present author~\cite{JHF1}, using a similar approach
 to Ref.~\cite{Ignatowsky}, it was shown that only two weak, and, from a naive point-of-view,
   `evident' postulates, unrelated to classical electrodynamics or any other
      dynamical theory, suffice to derive the LT. This is true for events lying along
    the common $x$, $x'$ axis of two frames S and S' in which the latter frame moves with constant
    velocity along the $x$-axis in S. The symbol $t$ is used to denote the recorded time (epoch)
    of a clock a rest in S, and $t'$ to denote the epoch of a clock a rest in S'.
    An event in S is specified by ($x$,$t$), one in S' by ($x'$,$t'$), it being understood
    that that $y = y'= z = z'=0$. The two postulates are that,
 the LT must be a single-valued function of its arguments, and a restricted form of the special
 relativity principle that may be called the `Measurement Reciprocity Postulate' (MRP):
   \par {\tt Reciprocal space-time measurements of similar measuring rods and clocks in two
   different inertial frames by observers at rest in these frames give identical
    results}
   \par In order to derive the LT for events not lying along the common $x$, $x' $ axis, the
    further postulate of spatial isotropy is required. This is, to the present writer's best
     knowledge, the minimum number of postulates, to date, from which the LT has been 
     derived~\cite{JHF1}\footnote{Einstein stated explicitly only two postulates in
     Ref.~\cite{Ein1}, The Special Relativity Principle and the constancy of the speed of light,
     but at least three other postulates: linearity of the equations,
      spatial isotropy, and the Reciprocity Principle (see below) were also tacitly
     assumed.} 
   \par The aim of the present letter is to present a simple thought experiment involving
    four identical unsynchronised  clocks A, B, A' and B' and a single measuring rod, R,
     that displays all essential predictions of SRT concerning
     measurements of length and time intervals in the inertial frames S and S'\footnote{The present paper
     supersedes the previous version~\cite{JHFAS}. The analysis of
     Ref.~\cite{JHFAS} although correctly concluding that the relativity of simultaneity and
     length contraction effects are spurious, was invalidated by an important conceptual error.
   This was conflating of kinematical configurations in a primary experiment and its reciprocal 
     with those in the frames S and S' in the primary experiment. See Refs.~\cite{JHFSTP3,JHFRECP}
       for the definition of `primary' and `reciprocal' experiments.}. The results
    obtained below from the analysis of this thought experiment are the same as those presented in
    previous papers~\cite{JHFLLT,JHFCRCS,JHFACOORD,JHFUMC} by the present author. Some are quite different
     from those deduced, following Einstein, from the conventional interpretation of
     the LT in SRT. Although the experimentally-confirmed Time Dilatation (TD) effect is
     predicted in the same way as in conventional SRT, neither the Relativity of
     Simultaneity (RS) nor the Length Contraction (LC) effects are predicted to occur.
      Indeed, the distance between two physical objects at rest in S' as measured in either
     S or S' is found to be a Lorentz-invariant quantity.
    It is interesting to note that, at
    the time of this writing, there is no experimental evidence for the existence of either
     the RS or LC effects~\cite{JHFLLT}. Two Earth-satellite experiments have recently
    been proposed by the present author to test for the existence of the RS effect~\cite{JHFSEXPS}.
     \par In order to specify a length interval, two distinct physical objects
    or a single extended one are required. For the present study it is found more 
    appropriate to consider two distinct objects, which may be chosen to be 
    identical clocks at two different spatial locations. This choice will be found to be of
    importance for discussion of the existence (or not) of the RS effect.
     Since the analysis will involve length intervals in the two inertial frames S and S',
      it follows that the minimum number of distinct objects needed in the thought experiment
      is four, ---two A,B at rest in S and
      two A',B' at rest in S'. The fixed separation of A and B (A'and B') specifies a length interval in S (S').
       In order to discuss TD and RS it will be convenient to use similar unsynchronised clocks for the four
     objects A, B, A' and B', although for the analysis of length intervals as measured
    in S or S' (relevant to the existence, or not, of LC) it is of no importance that
    the objects are clocks. Any others --- located at the same well-defined spatial
    locations--- can be employed. 
    \par As a first step, the relative separation of just two objects A and A' as viewed 
      in different inertial frames is considered. A' is placed at the origin of
    coordinates in the inertial frame S' so that $x'_{A'} = 0$, and is intially at the origin of S: $x_{A'}(t=0) = 0$.
    The object A, permanently at rest in S, is placed near
   the $x$-axis at:
          \[ x  = x_A = x_A- x_{A'}(t=0) \equiv  x_{AA'}(t=0) \equiv D  \]
    The object A' undergoes a time-dependent  proper acceleration, $\gamma(t')a_0$ (where  $a_0$ is constant,
    $\gamma(t') \equiv 1/\sqrt{1-(v_{A'}(t')/c)^2}$ and $v_{A'} \equiv dx_{A'}/dt$) along the positive $x$-axis
    during the time interval $t_{acc}$ in S, until it arrives at a position with the same
    $x$-coordinate as A. The corresponding proper time interval for A' is $t'_{acc}$. 
    The world line of A', as viewed from S,
       is given by the equation, derived in the Appendix\footnote{Note that the world line of A' corresponds
      to the `hyperbolic motion' that has formerly been associated with a constant proper acceleration.
      Why this is incorrect is explained in the Appendix.}:
      \begin{equation}  
       x_{AA'}(t) \equiv x_A- x_{A'}(t) = D-\frac{c^2}{a_0}\left[\sqrt{1+\left(\frac{a_0t}{c}\right)^2}-1\right]
      \end{equation}
    where $c$ is the speed of light in free space.
     Since $ x_{AA'}(t_{acc}) = 0$, it follows from (1) that:
  \begin{equation}  
       t_{acc} = \frac{c}{a_0}\sqrt{\left(1+\frac{a_0D}{c^2}\right)^2-1}
   \end{equation}
    As shown in Figs. 1a and 1b, the equation
     \begin{equation}
       x'_{AA'}(t') \equiv  x'_A(t') -  x'_{A'} = x_{AA'}(t) = D-x_{A'}(t)
    \end{equation}
     holds when (see the Appendix) $t$ and $t'$ are related according to the 
     equations:      
       \begin{equation}
      t'(t) = \frac{c}{a_0}\ln\left[\frac{a_0t}{c}+\sqrt{1+\left(\frac{a_0t}{c}\right)^2}\right]
            = \frac{c}{a_0}{\rm arcsinh}\frac{a_0t}{c},~~~ t(t') =  \frac{c}{a_0}{\rm sinh}\frac{a_0t'}{c} 
  \end{equation}
    Then (2) and the first equation in (4) gives:
       \begin{equation}
  t'_{acc} =  \frac{c}{a_0}{\rm arccosh}(1 +\frac{a_0D}{c^2})
  \end{equation}
      The world line of A as viewed from the proper frame of A', derived by combining Eqns(1),(3) and the second equation
    in (4) is:
   \begin{eqnarray} 
     x'_A(t') & = & x_{AA'}(t(t')) = D-\frac{c^2}{a_0}\left[\sqrt{1+\left(\frac{a_0t(t')}{c}\right)^2}-1\right] \nonumber  \\
              & = & D-\frac{c^2}{a_0}\left[ {\rm cosh}\frac{a_0t'}{c}-1\right]
 \end{eqnarray} 
    Since $ x'_A(t'_{acc}) = x'_{AA'}(t'_{acc}) = 0$ it follows from (6) that
   \begin{eqnarray} 
 t'_{acc} & = & \frac{c}{a_0}\ln\left[1 + \frac{a_0D}{c^2}+\sqrt{\left(1+\frac{a_0D}{c^2}\right)^2-1}\right]
    \nonumber \\
  & = &  \frac{c}{a_0}{\rm arccosh}\left(1+\frac{a_0D}{c^2}\right)
 \end{eqnarray}
    consistent with Eqn(5) above, thus verifying the correctness of Eqns(3) and (6).  
    The world line of A' as viewed from S, given by Eqn(1) is shown in Fig. 1a, and the world line
    of A as viewed from the proper frame of A', given by Eqn(6), is shown in Fig. 1b, for $a_0 = c = 1$
    and $t_{acc} = \sqrt{3}$   
    \par Differentiating (6) gives:
       \begin{equation}
 v'_{A}(t') \equiv \frac{d x'_A}{d t'} = -c~{\rm sinh} \frac{a_0t'}{c} = -a_0t = -v_{A'}(t) \gamma(t)
  \end{equation} 
    In deriving (8), the second equation in (4) and the relation (see the Appendix)
    \begin{equation}
   v_{A'}(t) \equiv \frac{d x_{A'}}{d t} = \frac{a_0t}{\sqrt{1+\left(\frac{a_0t}{c}\right)^2}} = \frac{a_0t}{\gamma(t)}
    \end{equation} 
    has been used. The slopes of the world lines in Fig. 1a and Fig. 1b at the corresponding
     values of $t$ and $t'$ given by Eqns(4) are therefore related by the Lorentz factor $\gamma(t)$.
     Thus the magnitude of the relative velocity of A and A'
    in the proper frame of A' is larger than than that in the inertial frame S, in which A 
    is at rest, by the factor  $\gamma(t)$. An identical relation holds between the 
     relative velocity of the two objects in S and that in the co-moving inertial frame
     of A'. If therefore the acceleration is halted at the corresponding epochs
      $t = t_{acc}$, $t' = t'_{acc}$ the proper frame of A' at this instant becomes 
      an inertial proper frame at all later times, so that subsequent time intervals
      $\Delta t$ and $\Delta t'$ satisfy the TD relation: 
      $\Delta t = \gamma(t_{acc}) \Delta t'$. Indeed just this relation is the basis of the
      derivation in the Appendix of the formulae (1), (4) and (9) above. 
       \par Since at any time after the acceleration program is halted,
  \begin{equation}
 \frac{\Delta x'_A}{\Delta t'} =  -\gamma(t_{acc})\frac{\Delta x_{A'}}{\Delta t}
   \end{equation} 
     where
 \[ \Delta x'_A \equiv  x'_A(t'+\Delta t')-  x'_A(t'),
   ~~~\Delta x_{A'} \equiv  x_{A'}(t+\Delta t)- x_{A'}(t) \]  
   the TD relation requires that
 \begin{equation}
     \Delta x'_A = -\Delta x_{A'} 
 \end{equation}  
   demonstrating the Lorentz invariance of corresponding length intervals in two
   inertial frames. Since both $x'_{A'}$ and $D$ are constant, the relation (11) is already
    implicit in Eqn(3) above. This behaviour is illustrated in Fig. 1c and 1d 
   where it is supposed that an acceleration procedure with a large value of $a_0$
  is applied to A' so that it obtains the final velocity $v_{A'}(t_{acc})$
    of Fig. 1a in a negligibly short interval of time. The proper frame of A' is
   subsequently the inertial frame S' and the world lines of A' and A are straight lines
    with gradients equal to those of the world lines in Fig. 1a and 1b at the times
    $t_{acc}$ and $t'_{acc}$ respectively. It is already clear from inspection
    of these plots and Eqns(10) and (11) that the physical basis of TD is
    not `length contraction', as in the conventional interpretation, but rather
    the different relative velocities of A and A' in the frames S and S' according to
    Eqn(8). 

\begin{figure}[htbp]
\begin{center}\hspace*{-0.5cm}\mbox{
\epsfysize15.0cm\epsffile{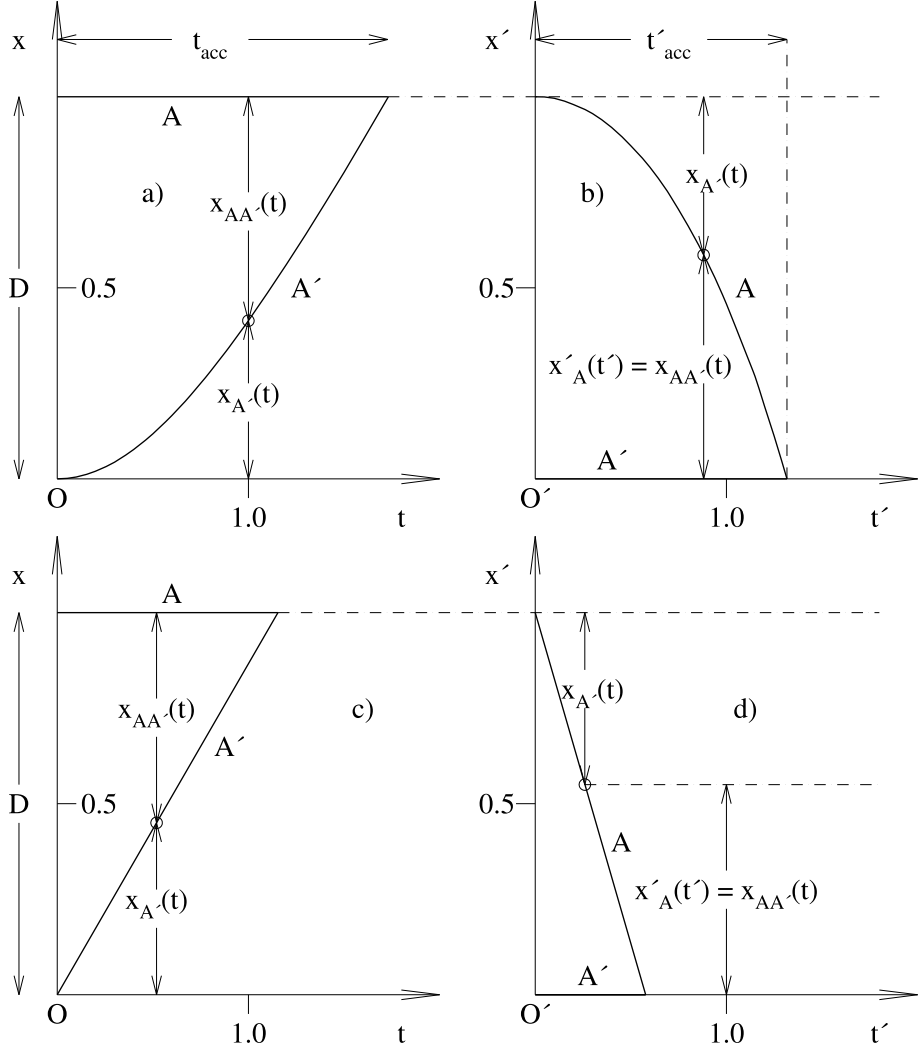}}
\caption{{\em  The world lines of the objects A and A'as viewed from the proper
  frame of A [ a) and c) ] or that of A' [ b) and d) ]. In a) and b) A' is accelerated
   for a time interval $t_{acc}$  in the inertial proper frame, S,
   of A. Units and dimensions are chosen so that $a_0 = c = D = 1$ and $t_{acc} = \sqrt{3}$. 
   In c) and d) a large value of $a$ is assumed so that A' reaches the same velocity 
   as in a) at  $t = t_{acc}= \sqrt{3}$ in a negligibly short time. The slope of the world line of A'
   in c) is then the same as the slope of the world line of A' in a) at $t = t_{acc}$.
    Similarly the slope of the world line of A in d) is equal to that of the world line of A
    in b) at  $t' = t'_{acc}= 1.317... $ The slopes of the world lines of A'[A] in c) [d)]
    are the same as those of the objects A',B' [A,B] in Fig. 3c [Fig. 3b] during the phase of
    uniform motion. As given by Eqn(8), the slopes of the world lines of A in b) and d)
    at corresponding epochs are $-\gamma(t)$ times the slopes of the world lines in a) and c) respectively.}}
\label{fig-fig1}
\end{center}
\end{figure}

\begin{figure}[htbp]
\begin{center}\hspace*{-0.5cm}\mbox{
\epsfysize15.0cm\epsffile{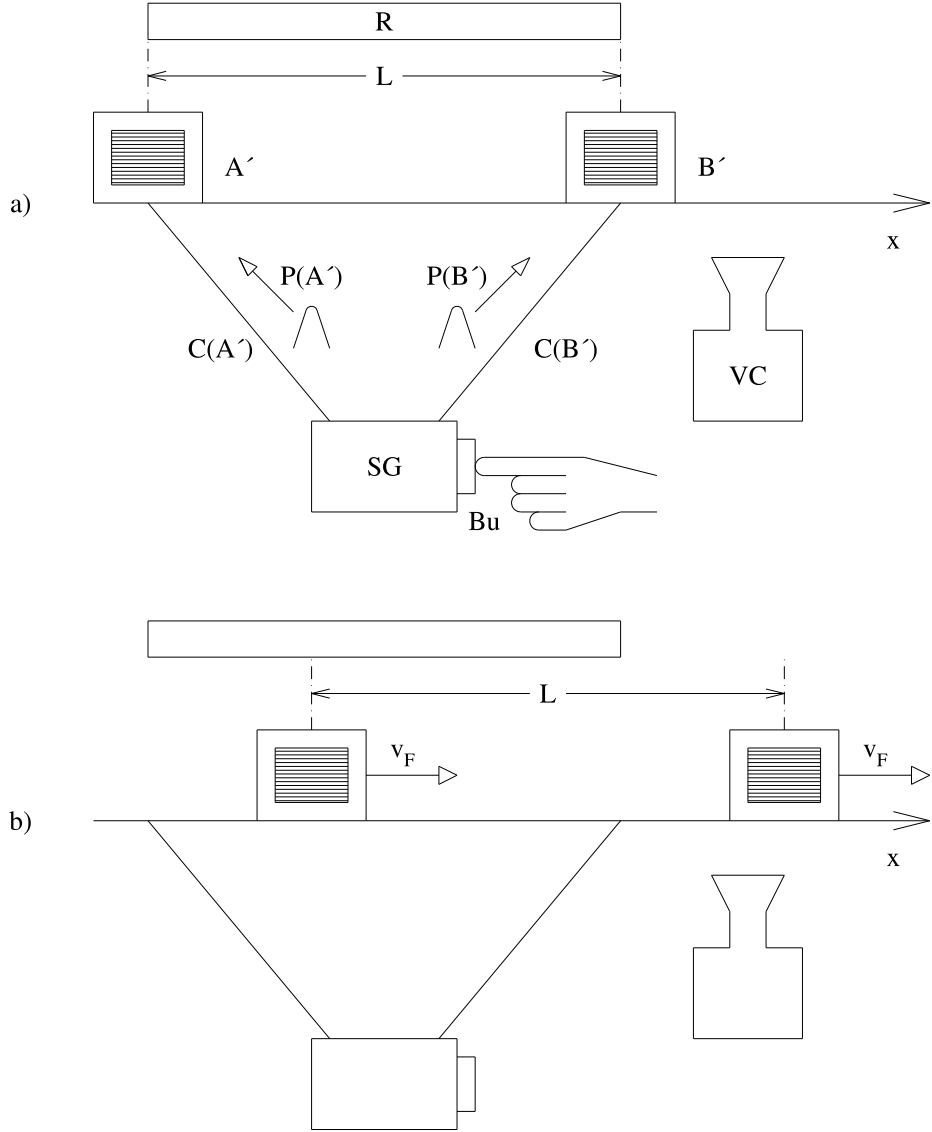}}
\caption{{\em a) Initial configuration of the clocks A' and B'
 at rest in the frame S. The measuring rod, R, of length $L$ is used to set the initial separation
 of the clocks. On pressing the button Bu, the signal generator, SG, sends simultaneous pulse-signals
 P(A') and P(B') along the identical cables  C(A') and C(B') to A' and B', respectively.
 On arriving at the clocks, these signals initiate identical acceleration programmes at epochs $t = 0$ for the clocks
 as described in Eqns(1) and (9). b) After the time interval $t = t_{acc} = \sqrt{3}$ (in units where $c = a_0 = 1$), the
  clocks move with the uniform velocity $v_F = \sqrt{3}/2$ along the $x$-axis, separated by the
   distance $L$. The video camera, VC, records the passage of the clocks (see text for discussion).}}
\label{fig-fig2}
\end{center}
\end{figure}

\begin{figure}[htbp]
\begin{center}\hspace*{-0.5cm}\mbox{
\epsfysize22.0cm\epsffile{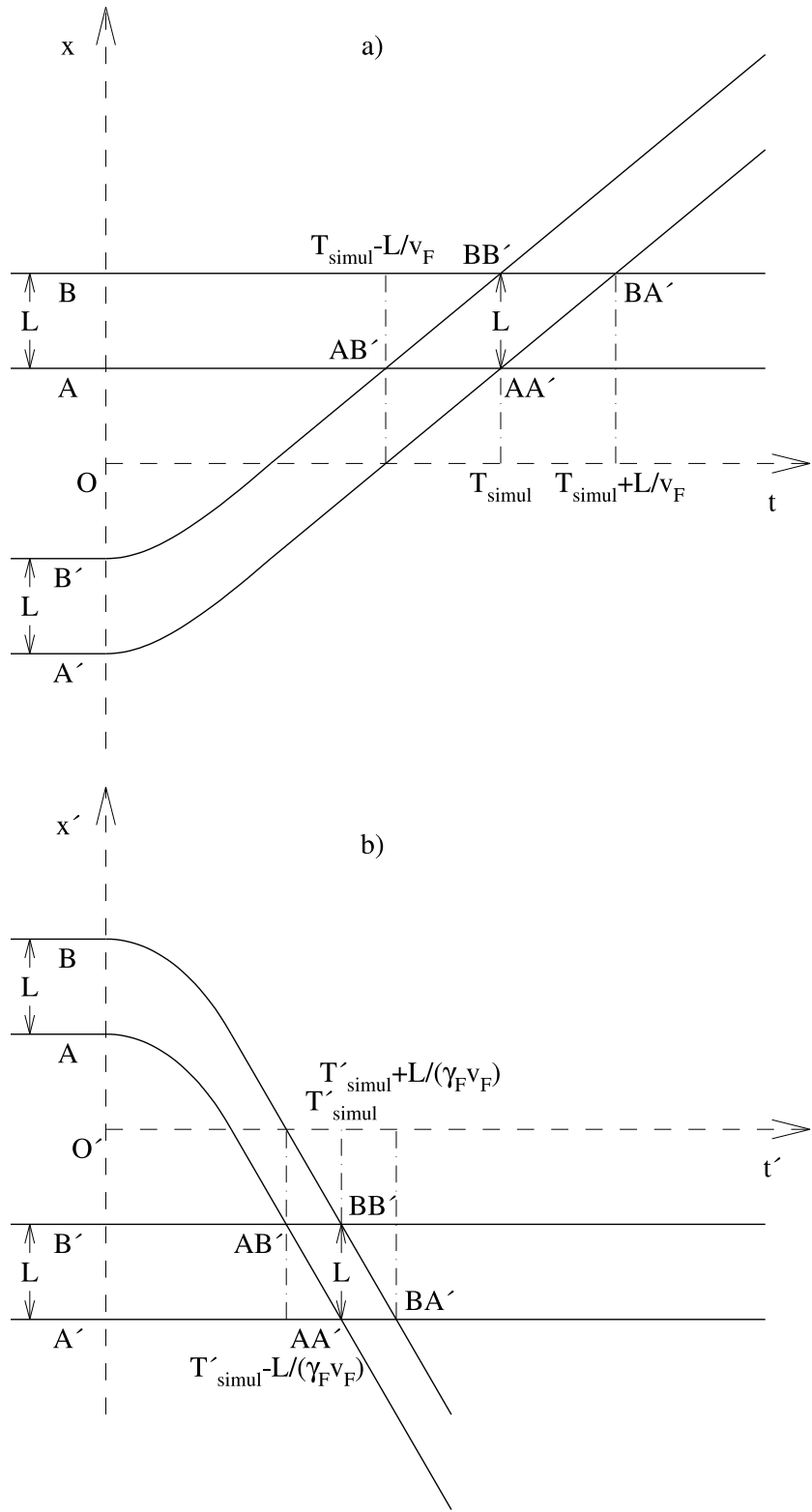}}
\caption{{\em  Space-time trajectories of the clocks as observed in: a), S and b), S'.
  S' is the instantaeneous co-moving frame of A' and B'. Units and dimensions with
   $a_0 = c = L = 1$ are used. At the epochs
   $t = T_{simul}= 4.15$,  $t' = T'_{simul}= 2.47$ clocks AA' and BB' are observed to be, simultaneously, in spatial coincidence
   in the $x$-direction, in both S and S'.}}
\label{fig-fig3}
\end{center}
\end{figure}

\begin{figure}[htbp]
\begin{center}\hspace*{-0.5cm}\mbox{
\epsfysize20.0cm\epsffile{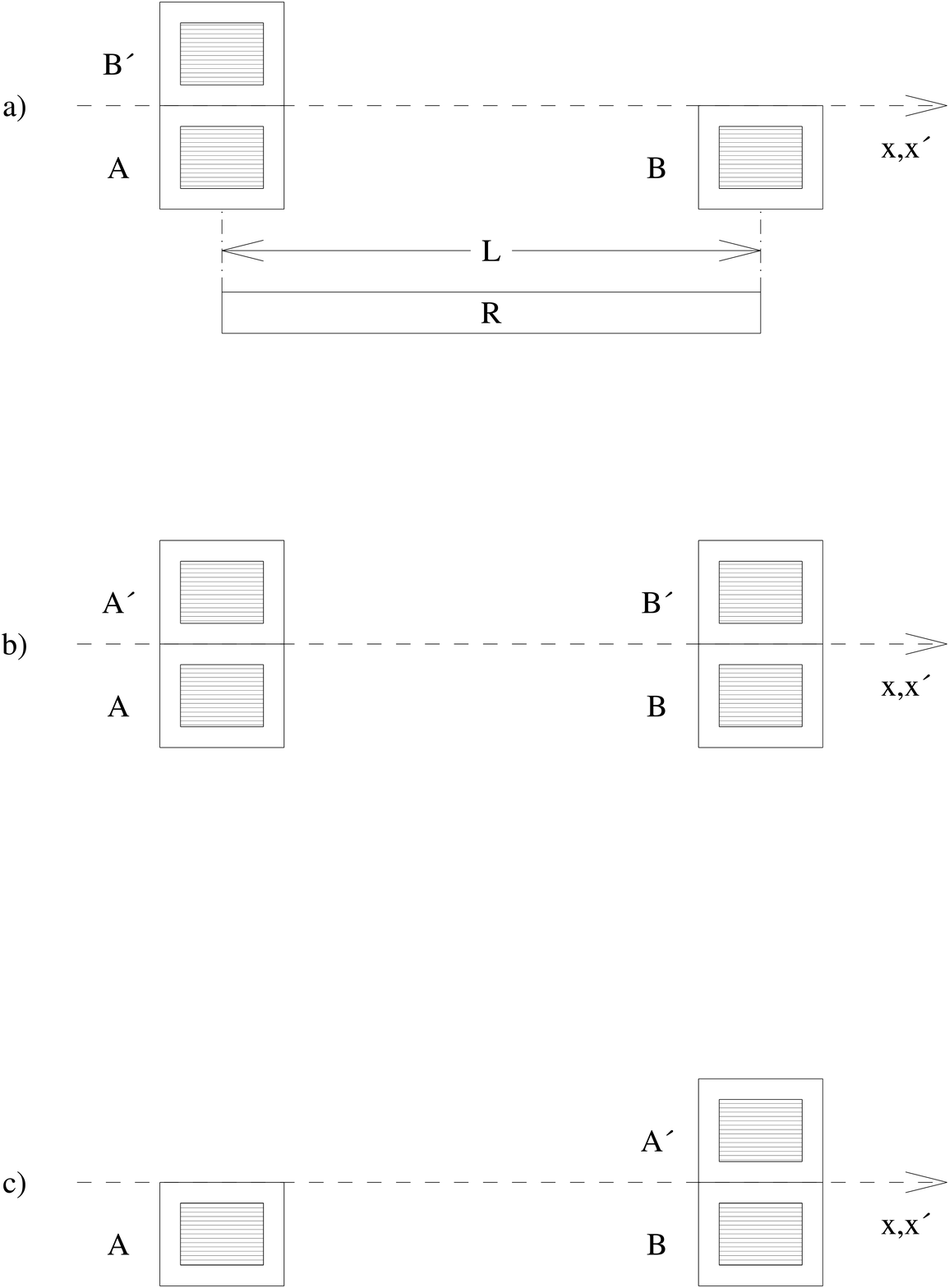}}
\caption{{\em Spatial configurations of the clocks as observed in either S or S' at different epochs. a)
  $t = T_{simul}-L/v_F$,  $t' = T'_{simul}-L/(\gamma_F v_F)$  b) $t = T_{simul}$, $t' = T'_{simul}$   c)
  $t  = T_{simul}+L/v_F$, $t'  = T'_{simul}+L/(\gamma_F v_F)$ .
   The spatial separation of A and B has been previously set to $L$, (see a)) using the measuring rod R.}}
\label{fig-fig4}
\end{center}
\end{figure}

    \par The thought experiment involving the four clocks A, A', B and B' and the measuring rod R
    will now be described. At the start of the experiment, as shown in Fig. 2a, the clocks A' and B'
     are placed in S
     above the $x$-axis, a distance, $L$, apart. The measuring rod, R, of length, $L$, is
      used to set the positions of the clocks. The clocks A and B are, similarly, set the same
    distance apart, using R, below the $x$-axis. Again, using R, the initial separation
    of B' and A is set to $2L$ and the origin of the $x$ and $x'$ axes is chosen midway between
    B' and A. Initially, then, the coordinates of the clocks are:
    $x_{A'} =x'_{A'} = -2L$, $x_{B'} =x'_{B'} = -L$,  $x_{A} =x'_{A} = L$  and
    $x_{B} =x'_{B} = 2L$. The clocks A' and B' are initially at rest but are equipped
 with identical  mechanisms that simultaneously accelerate them in the positive $x$-direction during the time interval
  $t_{acc}$ in the frame S. After this they move with the same uniform velocity $v_F$
  (see Fig. 2b) .
   It is therefore assumed that the same time-dependent acceleration program as discussed above, is applied
   simultaneously to both clocks in their common
   proper frame. The acceleration procedure is started simultaneously
   in S by signals from the generator SG which is connected via 
   the identical cables C(A') and  C(B') and movable contacts to A'and B'. 
  When the button Bu is pressed, simultaneous signal pulses P(A') and  P(B')
   propagate along the cables (Fig. 2a) and are detected by the accelerating mechanism
   in A' and B'. At time $t = 0$, according to a clock in S, both A' and B'
   start to accelerate in the direction of A and B. The velocities and positions of
   the clocks are given by the formulae (9) and (1) above, respectively, as:
  \begin{eqnarray}
             &   &~~ 0 < t \le t_{acc} \nonumber \\ 
   v_{A'}(t) & = &  v_{B'}(t) = \frac{a_0t}{\sqrt{1+\left(\frac{a_0t}{c}\right)^2}} \equiv v(t) \\
   x_{A'}(t)+2L & = &x_{B'}(t) +L = \frac{c^2}{a_0} \left[\sqrt{1+\left(\frac{a_0t}{c}\right)^2} -1\right]
   \end{eqnarray} 
  \begin{eqnarray}
             &   &~~~~~ t_{acc} < t   \nonumber \\ 
   v_{A'}(t) & = &  v_{B'}(t) = v(t_{acc}) \equiv v_F \\
   x_{A'}(t) & = & x_{B'}(t)-L = x_{A'}(t_{acc})+v_F(t-t_{acc})
   \end{eqnarray}
   The $x$ $v$ $t$ plot of the motion of A' and B' in S is shown in Fig. 3a. 
   At time $t = T_{simul}$, A,A' and B,B' are,
     simultaneously, spatially contiguous while at $t =  T_{simul} - L/v_F$ A,B' are contiguous
      and at
     $t =  T_{simul} + L/v_F$  B,A' are contiguous (see Fig. 3a and Fig. 4).
     \par The sequence of events as observed from the proper frame of of A'and B' is calculated
      with the aid of Eqns(6) and (7) above. These equations predict that the $x'$ versus $t'$ plots
    for A,B,A' and B' may be derived from those of Fig. 3a by reflection of the latter in the $x$ axis
      and by a suitable scaling of the slopes of the curves with different factors for the phases
      of accelerated and accelerated motion, the former scale factor, $-\gamma(t)$, being time-dependent
      and the latter, $-\gamma(t_{acc})$, constant. The space-time trajectories in Fig. 3b are then 
      given by the formulae:
 \begin{eqnarray}
             &   &~ 0 < t' \le t'_{acc} \nonumber \\ 
   v'_{A}(t') & = &  v'_{B}(t') =  -c~{\rm sinh} \frac{a_0t'}{c} \\
   L -x'_{A}(t') & = & 2L- x'_{B}(t') = \frac{c^2}{a_0}\left[ {\rm cosh}\frac{a_0t'}{c}-1\right] \\
   \end{eqnarray} 
  \begin{eqnarray}
             &   &~~~~~~ t'_{acc} < t'   \nonumber \\ 
   v'_{A}(t') & = &  v'_{B}(t') = v'_{A}(t'_{acc}) = -\gamma_F v_F \\
   x'_{A}(t') & = & x'_{B}(t')-L = x'_{A}(t'_{acc})-\gamma_F v_F(t'-t'_{acc})
   \end{eqnarray}
    \par The curves shown in Fig. 3 recall a phenomenon that the reader may have noticed when
     sitting in a train that has halted at a station. On the adjacent track are the carriages of
     another train. At a certain moment, the person in the train, looking out of the window,
     thinks that his train has started, as the carriages of the adjacent train are seen 
     to accelerate to the right. The last carriage passes and the person sees, with surprise,
     that his train is still stationary. Or, it may happen that instead the scenery outside
     is seen to be also moving to the right, at almost the same speed as the last carriage
     of the adjacent train, and the traveller is reassured that her journey is proceeding.
     In the first case the adjacent train accelerates to the right.
     In the second case the observer's train accelerates, in an identical manner, to the
     left. A physicist equipped with an accelerometer, could, of course, easily distinguish
     the two cases. Inspection of Fig. 1a and Fig. 1b shows that, in Special Relativity, in 
     contrast the Galilean relativity, if it is known that one of the two trains, but not 
      which one, has the acceleration program considered above, observation of the world line of
      one train from the other reveals which of them is accelerated, and which is
      at rest. If the shape of the world line of the observed train is described by Eqn(1),
      as in Fig. 1a, it is being accelerated, whereas, if it is described by Eqn(6) as in
       Fig. 1b, it is the observer's train which is undergoing acceleration. Since the functional
     dependence $\simeq \sqrt{1+(a_0t/c)^2}$ of Eqn(1) differs from that $\simeq  {\rm cosh}(a_0t/c)$
     of Eqn(6) careful observation of the times of passage of the ends of the carriges of the adjacent
     train can distinguish the two cases.
     \par Note that:
  \begin{eqnarray}        
 \frac{c^2}{a_ 0}\left[\sqrt{1+\left(\frac{a_0t}{c}\right)^2}-1 \right] & = & \frac{1}{2}a_0t^2
   - \frac{(a_0t)^4}{8c^2}+...  \\
 \frac{c^2}{a_ 0}\left[{\rm cosh}\frac{a_0t}{c}-1 \right] & = & 
 \frac{c^2}{2a_0}\left[\left(\exp\left[\frac{a_0t}{c}\right]+\exp\left[-\frac{a_0t}{c}\right]
 \right)-1\right] \nonumber \\ 
\nonumber \\ 
     & = & \frac{1}{2}a_0t^2 
       + \frac{(a_0t)^4}{24c^2}+... 
   \end{eqnarray}
   so that the functional dependence differs only by
   terms of O($(a_0t)^4/c^2$) and higher, which vanish in the Galilean limit where
   $c \rightarrow \infty$. In Fig. 2, the video camera VC records the motion
    of the objects A'and B'as they accelerate to the right and then move with uniform speed $v_F$.
     The motion of a particular point on the train follows the time dependence of Eqn(1) and (9)
     during the acceleration phase. If the VC were to be accelerated to the left in an identical manner,
     the motion of the same point on the train is instead described by Eqns(6) and (8), after reversing the
     direction of the $x'$-axis. In Galilean
      relativity, where only the $a_0t^2/2$ terms in (21) and (22) are retained the VC record is 
     identical in the two cases.

\begin{figure}[htbp]
\begin{center}\hspace*{-0.5cm}\mbox{
\epsfysize20.0cm\epsffile{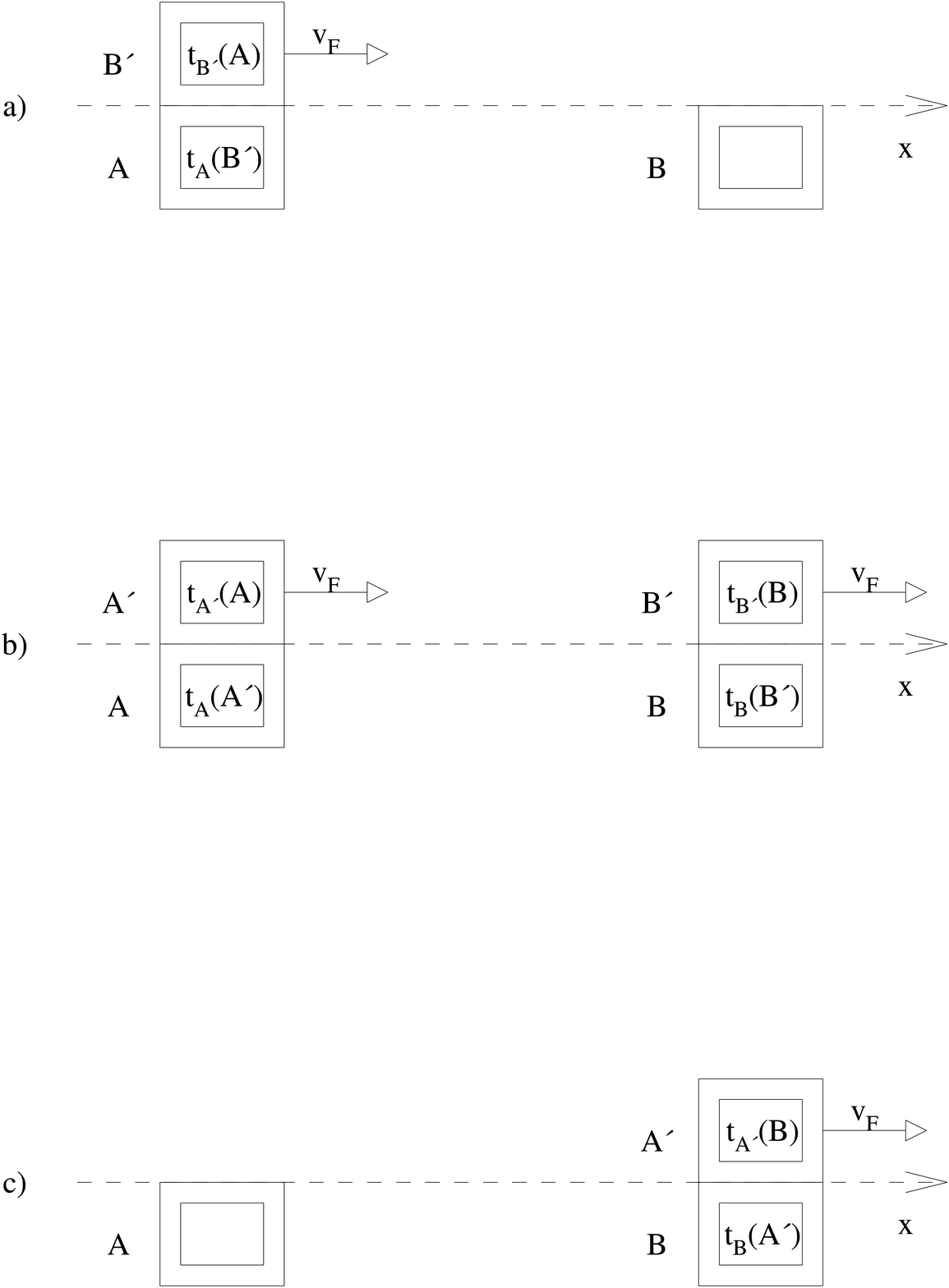}}
\caption[]{{\em Observations of the clocks from the frame S at different epochs:
 a) $t =  T_{simul}-L/ v_F$, b) $t = T_{simul}$, c) $t = T_{simul}+L/ v_F$.}}
\label{fig-fig5}
\end{center}
\end{figure}

\begin{figure}[htbp]
\begin{center}\hspace*{-0.5cm}\mbox{
\epsfysize20.0cm\epsffile{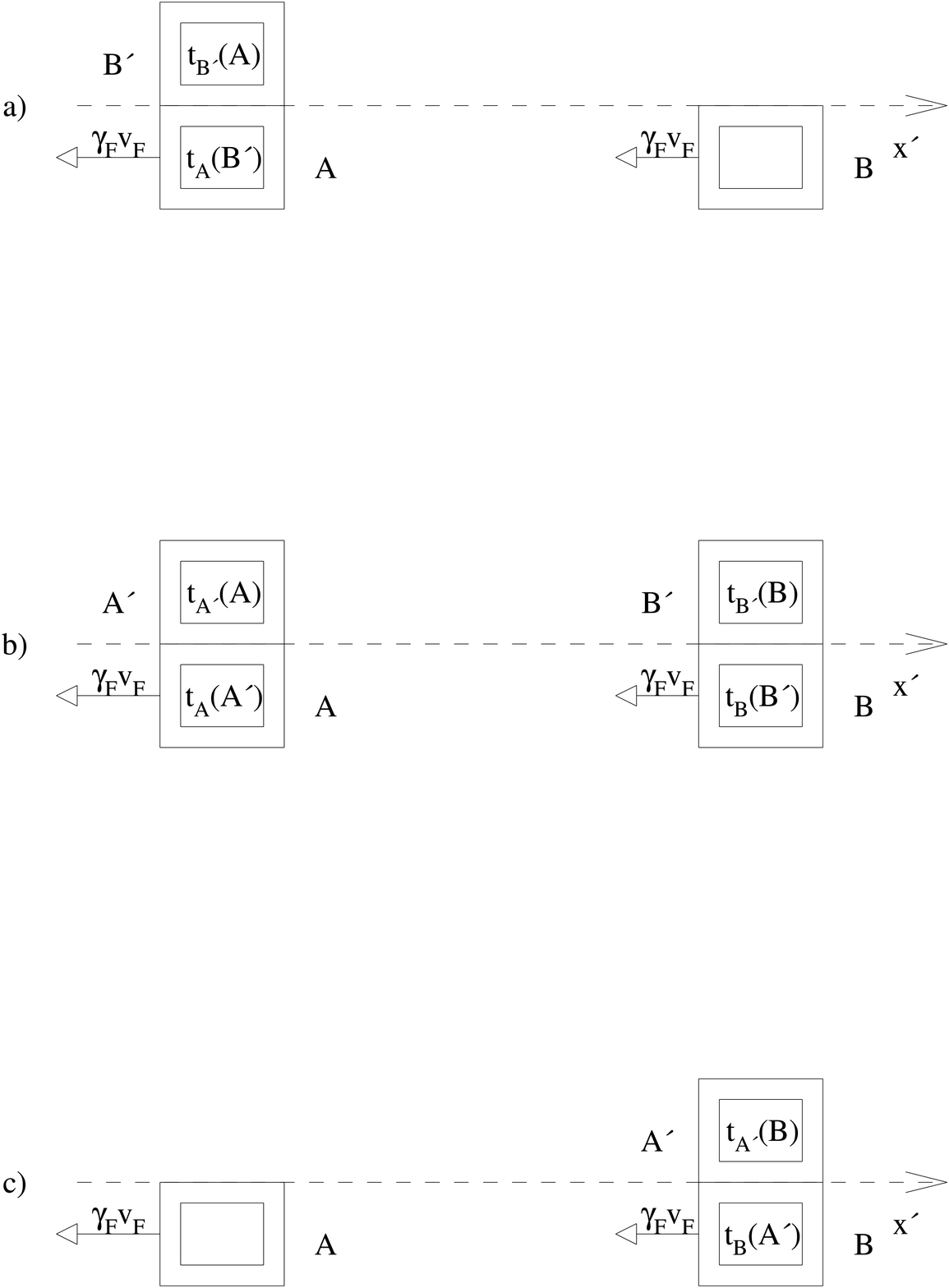}}
\caption{{\em Observations of the clocks from the frame S' at different epochs:
 a) $t' =  T'_{simul}-L/(\gamma_F v_F)$, b) $t' = T'_{simul}$, c) $t' = T'_{simul}+L/(\gamma_F v_F)$. }}
\label{fig-fig6}
\end{center}
\end{figure}
 
       \par The times recorded by the moving clocks A' and B', as viewed from the inertial frames S or S´ 
       during the phase of uniform motion, will now
       be considered. It is assumed that they are identical, unsynchronised, digital clocks that display
       a periodically updated sequence of numbers which are identified with the epoch $t_i$,
      for the clock $i$.  The times recorded by the clocks, at the events
      labelled in Figs. 3 and 4 by the proper times
        $T_{simul}-L/v_F$,  $T_{simul}$ and  $T_{simul}+L/v_F$ in S and
   $T'_{simul}-L/\gamma_F v_F$,  $T'_{simul}$ and  $T'_{simul}+L/\gamma_F v_F$ in S'
      are shown in Fig. 5, as observed in S, and in Fig. 6 as observed in S'. 
       The epochs are: $t_A(B')$, $t_A(A')$, $t_B(B')$, $t_B(A')$, $t_{B'}(A)$, $t_{A'}(A)$,
        $t_{B'}(B)$ and $t_{A'}(B)$. Here $t_i(j)$ is the epoch 
       recorded by the clock $i$ when it is in spatial coincidence, in the $x$-direction, with the
      clock $j$.  These eight observed epochs constitute the raw data of the thought experiment
      that will now be analysed using SRT.
       \par Consider first the observation of the clock B' between the AB' and
       BB' spatial coincidences. In the frame S (see Fig. 5a and 5b):
       \begin{equation}
         \Delta t(B',S) \equiv t_A(A')-  t_A(B') = \frac{L}{v_F}
       \end{equation}
        while in S' (see Fig. 6a and 6b)
  \begin{equation}
        \Delta t(B',S') \equiv t_{B'}(B)-  t_{B'}(A) = \frac{L}{\gamma_F v_F}
       \end{equation}
      combining (23) and (24) gives
 \begin{equation}
      \Delta t(B',S) = \gamma_F \Delta t(B',S')
 \end{equation}
   which is just the TD relation for the moving clock B'. Similarly for A':
    \begin{equation}
        \Delta t(A',S) \equiv t_B(A')-  t_B(B') = \frac{L}{v_F}
       \end{equation}
   and 
    \begin{equation}
         \Delta t(A',S') \equiv t'_{A'}(B)-  t'_{A'}(A)  = \frac{L}{\gamma_F v_F}
       \end{equation}
   which give the TD relation for the moving clock A':
    \begin{equation}
       \Delta t(A',S) = \gamma_F \Delta t(A',S')
 \end{equation}
   The TD effect for each of the moving clocks A'and B' involves two time intervals,
   one, $\Delta t(A',S')$  or  $\Delta t(B',S')$, recorded by the clock itself
     and the other,  $\Delta t(A',S)$  or  $\Delta t(B',S)$, recorded by a clock at rest in 
     S. For $\Delta t(A',S)$ (see Eqn(26)) this clock is B, whereas for $\Delta t(B',S)$,
      (see Eqn(23)) this clock is A. The above analysis makes very clear the physical
     origin of the TD effect in the larger relative velocity of A or B relative to
     A' or B' in the frame S' than in the frame S, the spatial separation between A'and B'
     and between A and B being equal at all times in both S and S'. 
    \par It has been assumed above that observers exist in both S and S' in order to note the 
     clock readings at the epochs of the AB', BB', AA' and BA' coincidence events as shown in 
      Figs. 5 and 6. All times intervals appearing in the TD relations (25) and (28) may
     thus be identified with observations of clocks at rest by two different observers ---one in 
     frame S, another in the frame S'. However, in all actual experimental realisations of the TD
     effect performed to date ---for example observation of the decay lifetime of an unstable
      elementary particle, $P$, in motion--- all observations are performed in the laboratory frame S.
      As in the above example,  $\Delta t(P,S)$ is still actually, or effectively\footnote{In the case of the
     muon decay experiment described in Ref.~\cite{MUDEC} the time interval is really
     measured by a clock in the laboratory frame. In more typical experiments~\cite{PIDEC} involving
    unstable particles with shorter lifetimes, it is deduced from the observed path length
    of the particle between production and decay and the velocity of the particle
       derived from kinematical measurements.}, recorded
      by a clock at rest in S, whereas  $\Delta t(P,S')$ is inferred from the measured values
       of $\Delta t(P,S)$ and $\gamma_P$. The TD relation is then verified by comparing the 
       calculated mean proper decay lifetime $\bar{\tau}$ of a statistical ensemble, $P_i$, $i =1,2,3...N$ of $N$ unstable 
       particles: 
         \begin{equation}
          \bar{\tau} = \frac{1}{N} \sum_i\frac{\Delta t(P_i,S)}{\gamma_{P_i}}
          \end{equation}
         with the previously measured mean lifetime
       for decay at rest for the same type of particle~\cite{PIDEC, MUDEC}.
        \par It is shown in Figs. 5 and 6 that the `Reciprocity Principle'(RP)~\cite{BG} that has hitherto
         been assumed to be valid in both Galilean and Special Relativity, does not hold in SRT.
         The RP states that: ``If the velocity of an object O' relative to an object O in the rest frame
         of O is $\vec{v}$ then the velocity of O relative to O' in the rest frame of O' is  $-\vec{v}$.''
          Eqn(8) above shows that this is not the case in SRT. As discussed at length in Refs.~\cite{JHFSTP3,
          JHFRECP}, although the RP does not apply to observations of the same space-time experiment
          in different frames, as, for example in Figs. 5 and 6 above, it does describe correctly the
           effect of a kinematical (velocity) LT between two inertial frames. For example the RP
          can be derived from the parallel velocity transformation relation (A.17) of the Appendix.
            In this case what are related are not events in the same experiment as viewed in 
            different inertial frames but rather the initial kinematical configurations of an experiment
            and its reciprocal as mentioned in the MRP above.
            If S' moves with velocity $v$ along the positive $x$-axis, in S, then setting
             $u = 0$ in (A.17), for the origin of S, gives $u'= -v$. In the primary experiment
             therefore the `travelling frame' S' moves with speed $v$ along the positive $x$-axis in
                the `base frame' S, whereas in the reciprocal experiment the travelling frame
               S moves with speed $v$ along the negative $x$-axis in the base frame S'.
              The RP thus relates base frame configurations in the primary and reciprocal experiments, not
              observed velocities in the frames S and S' of the primary experiment~\cite{JHFSTP3,JHFRECP}. 
             As shown for example, by the
           different nature of the TD effects ---the slower running clock of the primary experiment
          becomes the faster running one of the reciprocal experiment--- an experiment and its reciprocal
        are physically independent~\cite{JHFSTP3,JHFRECP}.
       \par Inspection of the geometry of Fig. 3 shows that the fundamental reason
        for the simultaneity in both S and S' of the AA' and BB' coincidence events,
        and the absence of any `length contraction' effect, is simply translational
        invariance (or the homogeneity of free space). The world line of B' is 
        obtained from that of A' in Fig. 3a, by the coordinate transformation $x \rightarrow x +L$,
       or that of B from that of A in Fig. 3b,
        by the transformation $x' \rightarrow x' +L$ ---the separation
        of B' from A' in Fig. 3a and of B from A in Fig. 3b is then independent of
       both their velocity and of time. The simultaneity of the events AA' and BB' in both S and S'
    is then a necessary consequence of the invariance of the separation of B from A 
     or B' from A' in both frames and of the equality of the two separations. All these
     consequences necessarily follow from the identical shapes of the world lines
    of A and B  and of A'and B' in both frames. This identity of form follows
    directly from the assumed identical acceleration programs of A' and B', independently
    of the nature of the program. The choice of `hyperbolic motion' considered
    above was made only for calculational convenience.
     \par The fallacious nature of the conventional predictions, of SRT, 
    following Ref.~\cite{Ein1}, of the RS and LC
       effects is explained, from different points of view, in Refs.~\cite{JHFLLT,JHFCRCS,JHFACOORD,JHFUMC}.
  \newpage
 \par {\bf Appendix}
\renewcommand{\theequation}{A.\arabic{equation}}
\setcounter{equation}{0}
 \par Suppose that  an object {\it O} moves with speed, $u({\it O})$ along the positive
  $x$-axis in the frame S, and that the object {\it O'} is at rest in the frame
      S' moving with (instantaneous) velocity  $v({\it O'})$ along the positive
  $x$-axis. The observed velocity of {\it O} along the positive $x'$-axis in S'
    is  then~\cite{JHFSTP3,JHFRECP}
   \begin{equation}
     u'({\it O}) =  \gamma_v[u({\it O})-v({\it O'})]
  \end{equation}
     where $\gamma_v \equiv 1/\sqrt{1-\beta_v^2}$, $\beta_v \equiv v({\it O'})/c$.
   If $u = 0$ at all times (i.e. for all values of $v({\it O'})$) then 
    \begin{equation}
      u'({\it O}) \equiv -v' =  -\gamma_v  v({\it O'}) = -c\gamma_v \beta_v 
  \end{equation}
   so that {\it O} is observed to move with speed $c\gamma_v \beta_v$ along the negative
      $x'$-axis in S'.
     Differentiating (A.2) with respect to the epoch, $t'$, recorded by the clock a rest in S',
      noting the differential TD relation:
       \begin{equation}
    dt = \gamma_v dt'
 \end{equation}
    gives
 \begin{equation}
      \frac{d v'}{dt'} = c\frac{d(\gamma_v \beta_v)}{dt}\frac{d t}{dt'} = c \gamma_v[1+(\gamma_v \beta_v)^2]
         \frac{d \beta_v}{dt}\gamma_v =  \gamma_v^4 \frac{d v({\it O'})}{dt} 
  \end{equation} 
   For the case of a constant proper acceleration of the frame S': $d v'/dt'$ = $a$ = constant, the equation
   given by transposing (A.4)
     \begin{equation}
       dt = \frac{c}{a} \frac{d \beta_v}{(1-\beta_v^2)^2}
  \end{equation}
    may be integrated, making use of the identity:
     \begin{equation}
            \frac{1}{(1-x)^2} \equiv \frac{1}{4}\left[\frac{2+x}{(1+x)^2} + \frac{2-x}{(1-x)^2}\right]
   \end{equation}
    to yield the relation between $t$ and $\beta_v$ for any object at rest in the accelerated frame
       S':
    \begin{equation}
     t(\beta_v) = \frac{c}{4a}\left[\frac{2\beta_v}{(1-\beta_v^2)} +\ln\frac{(1+\beta_v)}{(1-\beta_v)}\right]
   \end{equation} 
     The inverse of this equation $\beta_v(t)$ is therefore a transcendental function, preventing
     the determination, by analytical integration, of the equation of motion of {\it O'}.
     \par However, a simple analytical solution for the motion of {\it O'} may be obtained in the case
      that its proper acceleration is time-dependent:
      \begin{equation}
       \frac{d v'}{dt'} = a(t') = a_0 \gamma_v(t')
   \end{equation} 
     where  $a_0$ = $a(0)$ = constant. In this case (A.4) gives:
   
 \begin{equation}
      \frac{d v'}{dt'} = a_0 \gamma_v  =  \gamma_v^4 \frac{d v({\it O'})}{dt} 
  \end{equation} 
 so that 
      \begin{equation}
      a_0 = \gamma_v^3 \frac{d v({\it O'})}{dt}
  \end{equation}
  (A.10) may be integrated by noting the relation:
\begin{equation}
      d\left(\frac{\beta_v}{\sqrt{1-\beta_v^2}}\right) = \frac{d \beta_v}{(1-\beta_v^2)^\frac{3}{2}}
  \end{equation}    
      to yield, for an object initially at rest in the frame S:
\begin{equation}
    \frac{\beta_v}{\sqrt{1-\beta_v^2}} = \frac{a_0t}{c}
 \end{equation}
      or, transposing:
\begin{equation}
     \beta_v = \frac{1}{c} \frac{dx}{dt} = \frac{a_0t}{\sqrt{c^2 +a_0^2 t^2}}
 \end{equation}
       (A.13) is equivalent to Eqn(9) of the text.
      Integrating (A.13), assuming the same initial conditions as in (A.12) and (A.13) gives:
\begin{equation}
      x = \frac{c^2}{a_0}\left[\sqrt{1+\left(\frac{a_0t}{c}\right)^2}-1\right]
\end{equation}
      as used in Eqn(1) of the text.
      \par The relation between the epochs $t$ and $t'$ is given by integrating
       the TD relation:
       \begin{eqnarray} 
       t'(t) &  = & \int_0^t\frac{dt}{\gamma_v} =
          \int_0^t\frac{dt}{\sqrt{1+\left(\frac{a_0t}{c}\right)^2}} \nonumber \\
             & = & \frac{c}{a_0}\ln\left[at + \sqrt{1+ \left(\frac{a_0t}{c}\right)^2} \right]  \nonumber \\ 
             & = &  \frac{c}{a_0}{\rm arcsinh} \frac{a_0t}{c}
        \end{eqnarray}
     This is equivalent to Eqn(4) of the text. The relation $\gamma_v = \sqrt{1+ \left(\frac{a_0t}{c}\right)^2}$
     used in the third member of (A.15) is obtained by transposing (A.13) and using the definition of $\gamma_v$.
   The explicit $t'$ dependence of
    the acceleration program is then, using (A.15):
  \begin{equation}
   a(t') = a_0\gamma_v(t') =  a_0{\rm cosh} \frac{a_0t'}{c}
    \end{equation}
     \par The formulae (A.13) and (A.14) describing `hyperbolic motion' were first given by Born~\cite{Born}
     and Sommerfeld~\cite{Sommerfeld}. A formula equivalent to (A.10) was given
     by Rindler~\cite{Rindler},
     while Eqns(A.10),(A.13),(A.14) and (A.15) were all derived by Marder~\cite{Marder}.
    All of these authors associated
     these equations, not with the proper-time dependent proper acceleration of Eqn(A.16), but with
     a constant proper acceleration.
      The reason for these different predictions is that the analysis in the
      literature is based not on the relative-velocity transformation formula (A.2) but on the conventional
      parallel velocity addition formula of special relativity:
     \begin{equation}
     u' = \frac{u-v}{1-\frac{uv}{c^2}}
     \end{equation}
      where $u \equiv u({\it O})$, $u' \equiv u'({\it O})$ and $v \equiv v({\it O'})$ .
    Differentiating (A.17) with respect to $t'$ gives, with use of (A.3)
    \begin{eqnarray}
       \frac{d u'}{d t'} & = & -\gamma_v\left[\frac{1}{1-\frac{uv}{c^2}}-\frac{u}{c^2} 
        \frac{(u-v)}{(1-\frac{uv}{c^2})^2}\right]
            \frac{d v}{d t} \nonumber \\  
         &  & +\gamma_v\left[\frac{1}{1-\frac{uv}{c^2}}+\frac{v}{c^2} \frac{(u-v)}{(1-\frac{uv}{c^2})^2}\right]
            \frac{d u}{d t}
      \end{eqnarray}
     Since the condition $u = 0$ used to derive (A.2) from (A.1) holds at  all times, $du/dt = 0$. Setting then
  $u = 0$ and $du/dt = 0$ in (A.18) gives
 \begin{equation}
      \frac{d u'}{dt'} = -\gamma_v \frac{d v}{dt} 
  \end{equation} 
   which is inconsistent with (A.4) and does not agree with (A.10) when  $d u'/dt'$ = $a_0$ = constant.

    \par In order to derive (A.10) starting from (A.17), the procedure used in Refs.~\cite{Rindler,Marder}
     was firstly to hold $v$ constant in differentiating  (A.17) to give
  \begin{equation}
  \frac{d u'}{d t'} =  \gamma_v\left[\frac{1}{1-\frac{uv}{c^2}}+\frac{v}{c^2}
   \frac{(u-v)}{(1-\frac{uv}{c^2})^2}\right]\frac{d u}{d t}~~~( v = {\rm~constant} )
   \end{equation}
    and secondly to make the subsitution $u = v$ in this equation to give
 \begin{equation}
  \frac{d u'}{d t'} = \frac{\gamma_v}{1-\frac{v^2}{c^2}}\frac{d v}{d t}
     = \gamma_v^3\frac{d v}{d t}
   \end{equation}
    in agreement, up to a sign, with (A.10) when  $d u'/dt'$ = $a_0$ = constant. 
    There are two mathematical errors in the above derivation of (A.21):
    \begin{itemize}
    \item[(i)] Since this is a discussion where the frame S' is accelerated, $v$ is time-dependent, not constant.
     The postulate that $v$ is constant used to derive (A.20) from (A.18) therefore contradicts 
     an initial assumption of the problem.
   \item[(ii)] The condition $u = v$ implies that the object, the velocity of which is described by the left side of
    (A.21), is at rest in S' and moves with speed $v$ in the direction of the positive $x$-axis in S, whereas
    what must be described by this transformation equation (as is the case for (A.4)) is the velocity of
     an object at rest in S (so $u = 0$) as observed in the frame S'. In fact, since  $u \equiv u({\it O})$ represents
     the velocity of the object {\it O} whereas  $v \equiv v({\it O'})$ represents the velocity of the
     object {\it O'}, and the objects  {\it O} and  {\it O'} are distinct, the condition 
      $ u({\it O}) =  v({\it O'})$ implies that {\it O} and  {\it O'} (and hence S and S') are at rest
      relative to each other. The condition $u = v$ then also contradicts of the assumed initial
      condition of the problem that S' undergoes accelerated motion.
    \end{itemize}
      Apart from these purely mathematical errors, the use of the kinematical   
    velocity transformation formula (A.17) does not correctly describe the the velocity of {\it O}, as observed
     in the travelling frame S´ , in the space-time experiment in which {\it O'} has velocity
    $v$ in the base frame S. This is correctly given, not by (A.17), but by (A.2). The formula (A.17) instead
    transforms,
    between two inertial frames, the base frame kinematical configurations of two physically independent 
    space-time experiments, one primary and the other reciprocal. For further discussion of this essential
     conceptual point, see Refs.~\cite{JHFSTP3,JHFRECP}.

 \newpage


\end{document}